# Adaptive Video Encoding For Different Video Codecs

G. Esakki[1], A. S. Panayides[2,3], Senior Member, IEEE, V. Jalta[1], M. S. Pattichis[1], Senior Member, IEEE

[1]Department of Electrical and Computer Engineering, University of New Mexico, Albuquerque, NM 87131-0001 USA
[2]SiGiNT Solutions LTD, Nicosia 2311, Cyprus
[3]Department of Computer Science, University of Cyprus, Nicosia 1678, Cyprus

Corresponding author: A. S. Panayides (e-mail: panayides@cs.ucy.ac.cy; a.panayides@sigintsolutions.com).

This work was co-funded by the European Regional Development Fund and the Republic of Cyprus through the Research and Innovation Foundation (Project: POST-DOC/0916/0023); Acronym: ACTRESS.

**ABSTRACT** By 2022, we expect video traffic to reach 82% of the total internet traffic. Undoubtedly, the abundance of video-driven applications will likely lead internet video traffic percentage to a further increase in the near future, enabled by associate advances in video devices' capabilities. In response to this ever-growing demand, the Alliance for Open Media (AOM) and the Joint Video Experts Team (JVET) have demonstrated strong and renewed interest in developing new video codecs. In the fast-changing video codecs' landscape, there is thus, a genuine need to develop adaptive methods that can be universally applied to different codecs.

In this study, we formulate video encoding as a multi-objective optimization process where video quality (as a function of VMAF and PSNR), bitrate demands, and encoding rate (in encoded frames per second) are jointly optimized, going beyond the standard video encoding approaches that focus on rate control targeting specific bandwidths. More specifically, we create a dense video encoding space (offline) and then employ regression to generate forward prediction models for each one of the afore-described optimization objectives, using only Pareto-optimal points. We demonstrate our adaptive video encoding approach that leverages the generated forward prediction models that qualify for real-time adaptation using different codecs (e.g., SVT-AV1 and x265) for a variety of video datasets and resolutions. To motivate our approach and establish the promise for future fast VVC encoders, we also perform a comparative performance evaluation using both subjective and objective metrics and report on bitrate savings among all possible pairs between VVC, SVT-AV1, x265, and VP9 codecs.

**INDEX TERMS** Video codecs, Video signal processing, Video coding, Video compression, Video quality, Video streaming, Adaptive video streaming, Versatile Video Coding, AV1, HEVC.

## I. INTRODUCTION

Video streaming applications are steadily growing, driven by associate advances in video compression technologies. Video codecs support the widespread availability of digital video content that is dominating internet traffic compared to any other communicated data [1]. Traditional video sources such as video-on-demand (i.e., movies), teleconferencing, and live streaming events, including sports, concerts, and news, claim a significant share amongst the most popular applications, leveraging open-source protocols such as Dynamic Adaptive Streaming over HTTP (MPEG-DASH) [2]. Intriguingly, user-generated content facilitated by social media platforms has driven video communications to unprecedented levels [3]. Beyond that, synthesized video content augmented and virtual reality applications including 360º video content and point-cloud technologies [4]-[6], as well as internet gaming and emerging medical applications [7], [8] necessitate efficient solutions that will alleviate known bottlenecks, especially over resource constraint wireless networks.

Toward this direction, Versatile Video Coding (VVC)/ H.266 [9], the successor of the High Efficiency Video Coding (HEVC)/ H.265 standard [10], was officially released in July 2020 to reclaim the best compression efficiency available to the AV1 codec that was released in 2018 [11]. Both codecs target ultra-high-definition video coding, with a clear direction towards accommodating AR and VR applications, 360º and multi-view video coding.

VVC is the new compression standard of the Joint Video Experts Team (JVET), materialized by the long-lasting collaboration between the ISO/ IEC Moving Picture Experts Group (MPEG) and the ITU-T Video Coding Experts Group (VCEG). On the other hand, AV1 is the product of the Alliance for Open Media (AOM) [11], a new industry-driven initiative that has generated real competition with respect to the potential market penetration that a codec can have, not originating from the former two entities that have dominated the market over the past two decades.

Naturally, the afore-described developments generated a large body of literature in an attempt to document the coding capabilities of each codec and to demonstrate their advantages over legacy coders. The latter resulted in mixed results, mainly attributed to the fact that new and refined coding tools were gradually made available in their respective reference software implementations before their official releases. Recent studies appear to converge to the fact that VVC is the best codec available today, followed by AV1, which in turn outperforms HEVC [12], [13]. Here, it is worth highlighting that reference implementations of both HEVC/H.265 and H.264/AVC, namely HM and JM, respectively, perform better in terms of compression efficiency compared to popular open-source implementations, i.e., x265 and x264. On the other hand, the latter facilitates encoding optimizations that provide for real-time encoding capabilities, being orders of magnitude faster than the former and thus closer to industry adaptation.

A key contribution of the paper is to provide for a fair comparison of the recently released VVC standard [14] and AOM's debut AV1 encoder (using Intel's Scalable Video Technology SVT-AV1 implementation) [15], [16], along with both its predecessors, namely HEVC (x265 implementation [17]) and VP9 [18], respectively. More specifically, the paper provides bitrate comparisons in terms of subjective video quality assessment (VQA) [19], [20], the weighted Peak Signal to Noise Ratio ($PSNR_{611}$), and the Video Multi-Method Assessment Fusion (VMAF) metrics.

To support adaptive video encoding, we refer to the use of dynamic adaptive HTTP streaming (i.e., MPEG-DASH), also known as Hypertext Transfer Protocol (HTTP) adaptive streaming (HAS) that has proven instrumental in the success of popular industry video streaming platforms such as YouTube, Netflix, Apple, and Amazon. For this paper, we adopt the current practice of allowing the client to select video segment encodings that adapt to network conditions. To select a particular video segment encoding, the streaming server must have video segments pre-encoded at different quality levels and video resolutions. Per-title optimization (i.e., per video sequence) is a popular approach introduced by Netflix for adapting to different bitrate demands [21]. The challenge is then to perform an informed decision that will maximize the overall quality of the perceived video and thus user experience (QoE) [22]. Given the paradigm shift from RTP/RTCP pull-based approaches to HTTP/ Transmission Control Protocol (TCP) push-based ones in MPEG-DASH/ HAS, erroneously received video packet effects are mitigated. Hence, quality of service (QoS) parameters that are now in need to be optimized mostly concern minimizing (i) the initial time required to commence video rendering (i.e., time is taken to initialize video buffer) and (ii) buffering events (buffer becomes empty due to limited available bandwidth prohibiting the download of the next video segment) that result in a video freeze [22]. A significant body of literature exists studying efficient adaptation algorithms that can be broadly categorized with respect to the dominant adaptation parameter considered. Rate-based techniques optimize the utilization of the available bandwidth, while buffer-based ones are concerned with maintaining the best possible buffer fullness ratio. Recently, collaborative rate and buffer-based and quality-based (provided video chunk's quality is available) are widely used and studied methods [23]-[25].

The challenge when it comes to implementing such algorithms lies in the time-varying nature of (wireless) transmission mediums resulting in significant variations of the available bandwidth. The latter is especially true in view of the established user demands that have also been cultivated by the industry for anywhere, anytime, and any device, robust access to video services. Moreover, optimization objectives are competing in the sense that increased video quality translates to increased bitrate and complexity (i.e., for higher video resolution decoding). In addition, video applications are recourse hungry and can cause battery drainage in mobile devices.

In this study, we propose the use of a multi-objective optimization approach that jointly maximizes video quality while minimizing bitrate and encoding time demands. The approach is codec-agnostic and can thus be trained and used over different codecs in a multi-codec adaptive video encoding setting. The proposed methodology, subject to minor modifications, can be used for both server-side and/ or client-side adaptive video streaming applications. In the particular setup depicted in the present manuscript, server-side adaptive encoding for real-time video streaming scenarios is examined, utilizing client-side feedback concerning the instantaneous throughput to guide the adaptation process. In the typical MPEG-DASH setting, the proposed approach can be fine-tuned using the decoding time instead of the encoding time. The proposed approach significantly extends prior work by our group involving single-resolution medical videos encoded using the x265 video coding standard alone [26].

Key contributions of the present study include:
- Comprehensive video codecs performance evaluation using objective metrics (e.g., PSNR, VMAF), and subjective evaluations,
- Adaptive video encoding leveraging multi-objective optimization of key video parameters including quality, bitrate demands, and encoding rate,

TABLE I
VIDEO DATASETS CHARACTERISTICS

| Dataset Name | No. Videos | Video Resolution | Frame Rate | Duration |
|---|---|---|---|---|
| HEVC Dataset | 1 | 416x240 | 30 | 10s |
| HEVC Dataset | 2 | 416x240 | 50 | 10s |
| HEVC Dataset | 1 | 416x240 | 60 | 10s |
| UT Live Dataset | 7 | 768x432 | 25 | 10s |
| UT Live Dataset | 3 | 768x432 | 50 | 10s |
| HEVC Dataset | 2 | 832x480 | 30 | 10s |
| HEVC Dataset | 2 | 832x480 | 50 | 10s |
| HEVC Dataset | 6 | 1280x720 | 60 | 10s |
| HEVC Dataset | 2 | 1920x1080 | 24 | 10s |
| HEVC Dataset | 2 | 1920x1080 | 50 | 10s |
| HEVC Dataset | 1 | 1920x1080 | 60 | 10s |
| HEVC Dataset | 2 | 2560x1600 | 30 | 5s |
| **Total Videos** | **31** | | | |

- Cross-codec applicability across different datasets and video characteristics, and
- VMAF-driven adaptive video encoding.

The rest of the manuscript is organized as follows. Section II details the methodology for the video-codecs performance comparative evaluation as well as the multi-objective optimization setup for adaptive video encoding. Then, Section III discusses the results obtained for both scenarios. Finally, Section IV provides the concluding remarks and highlights ongoing and future work.

## II. METHODOLOGY

In the present section, we detail the methodology used to perform the comparison between the video codecs' compression efficiency. The latter involves the video datasets considered in this study, followed by the experimental setup and the video quality assessment (VQA) methods. Then, we describe the multi-objective optimization approach used to achieve adaptive video encoding to match time-varying available bandwidth using different compression standards based on the same video encoding and assessment methodology.

### A. MATERIAL – VIDEO DATASETS

For the purposes of this study, we use two established datasets: the HEVC test sequences [27] and the UT LIVE video dataset [28]. The former consists of video resolutions ranging from 240p to 1600p with corresponding frame rates ranging between 24 and 60 frames per second (fps), summing 21 videos of different content. The latter consists of ten videos of 768x432 video resolution at 25 fps (7 videos) and 50 fps (3 videos). Overall, 31 video sequences were used in the present study to deduct the compression capabilities of the examined video compression standards that are summarized in Table I. A representative sample of the incorporated video content appears in Fig. 1. A subset of these videos was further used to train and validate the multi-objective video adaptation framework.

### B. VIDEO CODECS PERFORMANCE EVALUATION
1) VIDEO ENCODING EXPERIMENTAL SETUP

The selected encoding parameters per video codec appear in Table II. Comparable encoding setups were selected based on state-of-the-art literature to allow for a fair comparison, as further detailed in previous studies by our group [12], [13]. As such, the comparative performance evaluation relied on constant quality encoding. Selected quantization parameters comprised of {27, 35, 46, 55} for SVT-AV1 and VP9, AOM/ Intel, and Google codecs, respectively, while matching QPs of {22, 27, 32, 37} for JVET's VVC and x265 (JCT-VC HEVC/H.265 open-source implementation) were used. To support random access for VVC, an intra update was inserted every 32 frames for 25fps and 40fps videos and every 48 frames for 50fps and 60 fps videos. For the remaining codecs, random access intervals were set at the video's frame rate. Bi-prediction (B/b frames in VVC and x265 and alternate reference frames in AV1 and VP9 codecs), enabling both preceding and subsequent pictures to be used for spatial and temporal prediction and compensation of the currently encoded frame, was employed. Additionally, codec-specific filters were enabled towards enhancing video fidelity. The remaining parameters were tuned according to predefined presets available in every codec as highlighted in Table II, namely –random access for VVC, default preset for SVT-AV1, --ultrafast for x265, and --rt for VP9.

2) VIDEO QUALITY ASSESMENT

Video quality assessment of encoded sequences relied on popular objective video quality metrics as well as perceptual (subjective) assessment of a representative subset of the compressed video instances as described next. Here, we use popular Peak-Signal-to-Noise-Ratio (PSNR) as the benchmark metric that will allow direct comparison to similar studies in the literature. However, due to the rather weak correlation of PSNR to perceptual quality, we further employ the state-of-the-art Video Multi-Method Assessment Fusion (VMAF) metric [29], [30]. VMAF has shown consistently high correlations to subjective (user) ratings across video resolutions and datasets, and hence VMAF scores provide for a more realistic representation of a video's quality [13], [29]. We also consider perceptual (subjective) video quality assessment to (i) to verify the afore-described hypothesis that VMAF achieves better correlation to perceptual scores than PSNR, and (ii) to further validate the performance of the video codecs.

*I. Objective Video Quality Assessment*
Peak-Signal-to-Noise-Ratio (PSNR) is considered the de-facto benchmark metric being used for decades in image and video quality assessment. To compensate for the limited

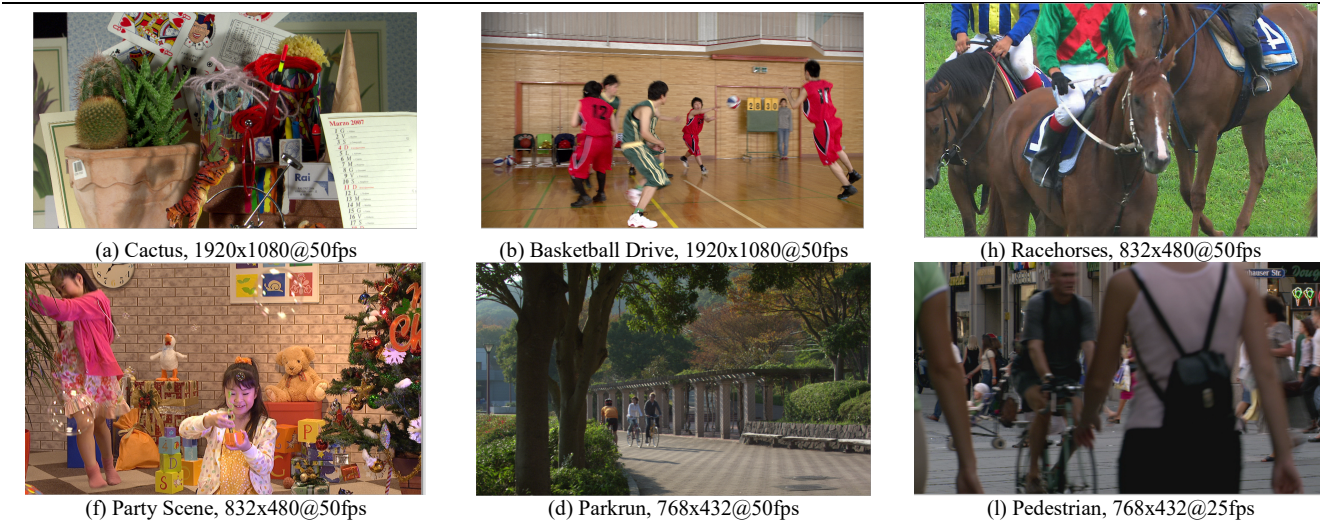

**FIGURE 1.** Video image examples of the investigated datasets capturing different content to provide for a realistic video codecs comparison. For each depicted video sequence, the name, the video resolution and the frames per seconds are displayed.

TABLE II
VIDEO CODECS EXPERIMENTAL SETUP

| Codec | Version | QP Range | Encoding Parameters |
|---|---|---|---|
| VVC | 7.1 | 22, 27, 32, 37 | --random access, --sao, --loopfilter |
| X265 | 2.1 | 22, 27, 32, 37 | --psnr, --ssim, --deblock, --ultrafast, --no-open-gop -F 4 -s 64, -b3 --key-int <framerate> |
| SVT-AV1 | 0.7 | 27, 35, 46, 55 | --enable-altrefs=1, --altref-nframes=2, --hierarchical-levels=4 -base-layer-switch-mode=1, --restoration-filtering=0, --dlf=0, --psnr, -intra-period=<framerate>, --arnr-maxframes=7, --arnr-strength=5, --arnr-type=3, --end-usage=3, --bit-depth=8 |
| VP9 | 1.8 | 27, 35, 46, 55 | --profile=0, --psnr, --cpu-used=8, --rt --tune=psnr, --arnr-maxframes=7, --arnr-strength=5, --arnr-type=3 --row-mt=1, --end-usage=3 , --bit-depth=8, --kf-max-dist= <framerate>, --auto-alt-ref=2, --lag-in-frames=25 |

[a]Random Acess was set at 32 for 25fps and 40fps videos, and 48 for 50fps videos, respectively.

correlation to the actual perceived quality, a new weighted alternative is employed nowadays:

$$PSNR_{611} = (6\ PSNR_Y + PSNR_U + PSNR_V)/8 \quad (1)$$

where $PSNR_Y$ refers to the luma component whereas $PSNR_U$ and $PSNR_V$ correspond to the chrominance ones. $PSNR_{611}$ considers the brightness intensity (luma) as the most decisive contributing factor of the raw YUV space driven by the observation that a human eye is more susceptible to brightness variations than colors. $PSNR_{611}$ was introduced during HEVC development [31] and was found to correlate better to subjective quality (with respect to traditional PSNR) while providing a combined score for luma and chrominance assessment, ideally suited for computing Bjontegaard Delta (BD-rate) differences [32].

Video Multi-Method Assessment Fusion (VMAF) [29], [30], as its name suggests, relies on the fusion of known image quality assessment metrics applied for VQA (Visual Information Fidelity (VIF) [33]; Detail Loss Metric (DLM) [34]) in conjunction with the computation of the motion offset between subsequent frames. Importantly, VMAF uses machine learning to train on a specific dataset and thus increase its correlation to subjective ratings. Since its introduction in 2016, it is one of the most widely used VQA methods in both industry and academia. Here, we use VMAF as the primary VQA algorithm (using the default, pre-trained machine learning coefficients, i.e., no training on the examined datasets has been performed) while $PSNR_{611}$ is employed for benchmark purposes.

Finally, to compute the bitrate gains and/ or reductions of one video codec over the other, the Bjontegaard Delta (BD) metric [32] was employed (see [12], [13], for more details).

## II. Subjective Video quality assessment

### Selected Video Instances

To validate the perceptual quality of the compressed video instances, a balanced subset in terms of video quality, video content, video resolution, and the frame rate was abstracted from the total number of the investigated videos in this study. More specifically, 12 videos originating from both the HEVC and the UT LIVE datasets, consisting of 416x240p, 768x432p, 832x480p, 1280x720p, 1920x1080p and 2500x1600p video resolutions were selected. For each of the aforementioned videos, three different rate points (quantization values) were used, categorized with respect to their objective VMAF scores as follows:

- High Quality: VMAF score ≥ 85
- Medium Quality: 60 ≤ VMAF score ≤ 70

TABLE III
NOTATIONS AND ACRONYMS DEFINITION

| Acronym/Notation | Description |
|---|---|
| VMAF | refers to Video Multimethod Assessment Fusion in the range of 0-100 |
| PSNR | refers to PEAK Signal to Noise Ration in dB |
| QP | refers to Quantization Parameter |
| GOP | refers to Group of Pictures, i.e., B2, B3, B4, B6, ZL in x265[a] |
| **Optimization Objectives and their Bounds** | |
| VQ | denotes Visual Quality. It can be PSNR or VMAF |
| $VQ_{min}$ | denotes the minimum quality bound. The video needs to be encoded in higher quality than this bound |
| B | denotes the encoded video bitrate demands |
| $B_{max}$ | denotes the maximum available bitrate. The video must be encoded below this bound |
| FPS | denotes the encoded video frames per second |
| $FPS_{min}$ | denotes the minimum bound for the requirement of how fast the video needs to be encoded |
| **Optimization Model Parameters** | |
| $i$ | is used to index the video segment $i$=1, 2, …. |
| $c$ | is used to index the video encoding configuration. Except for QP, $c$ refers to the video encoding configurations given in Table IV |
| $j$ | is used to index the objectives. $j = 1$ for VMAF, $j = 2$ for PSNR, $j = 3$ for B (Bitrate), $j = 4$ for FPS. |
| $k$ | is used to denote the power of the quantization parameter QP. $k = 0$ for $QP^0 = 1$. $k = 1$ for $QP^1 = QP$. $k = 2$ for $QP^2$. |
| $\beta_{i,c,j,k}$ | regression model coefficient for video segment= $i$, enc. conf.= $c$, objective= $j$, QP power= $k$. |

[a]See Table IV for different options

- Low Quality: VMAF score ≤ 50

The latter categorization translated and allowed the collection of three different video quality categories, namely low, medium, and high quality, across the investigated video resolutions. The objective was to capture a balanced and adequate sample of the compressed video instances that would allow a safe extrapolation of the perceptual scores over the whole dataset. In other words, to use a representative video sample that would highly approximate the outcomes of the perceptual evaluation if this were to be performed over the entire set of investigated videos. Here, it is important to highlight that subjectively evaluating the entire set of compressed video sequences was not feasible due to time constraints. Overall, a total of 108 video instances were assessed (12 videos x 3 QPs x 3 codecs) vs. a total of 496 encoded video instances used in this series of experiments (31 videos x 4 QPs x 4 codecs).

*Training and Testing Sessions*

The encoded videos were evaluated over three distinct sessions, each one corresponding to one of the examined encoders, namely SVT-AV1, x265, and VP9. Reliable VVC-encoded video rendering was not possible at the time of the experiments. The study involved 32 volunteers. Before initiating the actual scoring session, participants were given a scoring sheet with the following categories:

1-Bad, 2-Worse, 3-Fair, 4-Good, 5-Excellent

Each scoring entry was then explained using a video example of matching quality. The videos used for this purpose were not part of the perceptual VQA dataset. Then, the assessment procedure was described, and participants were given as much time as it was required to familiarize themselves with the involved processes before the examined test sequences were displayed in a randomized fashion.

All evaluations were performed using a SAMSUNG U28E690D LCD TV with a spatial resolution of 3840 x 2160mm and maximum screen brightness (peak luminance of 370 lux.) in a brightly lit room. Optimal 3H viewing distance was secured for all participants [35], [22].

### C. ADAPTIVE VIDEO ENCODING

An abstract system diagram of the proposed adaptive video encoding framework appears in Fig. 2. The method relies on a two-step process, of which the first step is performed offline and generates forward prediction models, while the second one capitalizes on these models to perform real-time adaptive video encoding. We start by producing a dense encoding space of the investigated video segments to capture different content characteristics found within the same video sequence. Then, Pareto front sampling is performed to identify non-dominated encoding setups, followed by curve fitting using regression to produce the forward prediction models. Forward prediction models are generated per an optimization objective, namely video quality, bitrate, and encoding rate (in frames per second - FPS). These models allow us to predict encoding configurations that match the time-varying network characteristics during real-time video streaming. In other words, a significant variation in available bandwidth (being an increase or decrease) will trigger the proposed adaptive video control mechanism (described below) to generate a new encoding configuration. For our proposed maximum video quality mode, the new encoding configuration will have to meet constraints on the currently available bitrate while securing the best possible video quality and an encoding rate for real-time streaming services.

Our proposed adaptive video coding is more responsive than typical encoding approaches used with MPEG-DASH, where pre-encoded video segments cover a sparser sampling of available bandwidths and associated video resolutions [21], [36]. As a result, we would expect the proposed methods to translate into fewer buffering and video stalling incidents. To this end, we propose the minimum bitrate mode that minimizes bitrate requirements while meeting or exceeding constraints on video quality and the encoding rate. For the proposed maximum performance mode, we emphasize real-time performance while meeting constraints on available bandwidth and required video quality. Overall, our proposed models and optimization methods provide strong adaptation to different scenarios.

### 1) VIDEO MODELING AND OPTIMIZATION

In this section, we will provide a detailed description of how we model the different objectives, compute the relevant regression models, and compute the optimal encoding parameters. We then demonstrate how to apply the models to optimize for maximum video quality, minimum bitrate, and maximum performance.

In order to model strong variations of video content, we break the video into short video segments of three seconds. Then, over each segment, we model the objectives as functions of the encoding configuration and the quantization parameter. We thus use a large number of models as functions of the video segment $i$ and the encoding configuration $c$. In what follows, we summarize the mathematical notation in Table III.

For each video segment $i$ and encoding configuration $c$, we model the objectives as functions of the quantization parameter given by (see Table III for sub-indices for $\beta_{i,c,j,k}$):

$$\begin{bmatrix} \log(\text{VMAF}) \\ \log(\text{PSNR}) \\ \log(B) \\ \log(\text{FPS}) \end{bmatrix} = \begin{bmatrix} \beta_{i,c,1,0} & \beta_{i,c,1,1} & \beta_{i,c,1,2} \\ \beta_{i,c,2,0} & \beta_{i,c,2,1} & \beta_{i,c,2,2} \\ \beta_{i,c,3,0} & \beta_{i,c,3,1} & \beta_{i,c,3,2} \\ \beta_{i,c,4,0} & \beta_{i,c,4,1} & \beta_{i,c,4,2} \end{bmatrix} \begin{bmatrix} 1 \\ \text{QP} \\ \text{QP}^2 \end{bmatrix}. \quad (2)$$

The regression models are estimated using least squares. To demonstrate the approach, let us consider the model for VMAF. In this example, we will be estimating the values for $\beta_{i,c,1,0}$, $\beta_{i,c,1,1}$, and $\beta_{i,c,1,2}$. First, we will need to encode the video by varying the quantization parameter and measuring the corresponding values for VMAF for the encoded videos. We can also measure PSNR, B, and FPS using the same video encodings. The procedure for estimating the regression models for PSNR, B, and FPS is similar.

For each video segment $i$ and encoding configuration $c$, let the video be encoded using $n$ values of the quantization parameter is given by:

$$\text{QP} = \text{QP}_1, \text{QP}_2, \ldots, \text{QP}_n.$$

Suppose that the corresponding VMAF values for each QP are given by:

$$\text{VMAF} = \text{VMAF}_1, \text{VMAF}_2, \ldots, \text{VMAF}_n.$$

We then formulate the regression model using:

$$y = X\beta + \varepsilon \quad (3)$$

where:

$$y = \begin{bmatrix} \log(\text{VMAF}_1) \\ \log(\text{VMAF}_2) \\ \vdots \\ \log(\text{VMAF}_n) \end{bmatrix}, \quad X = \begin{bmatrix} 1 & \text{QP}_1 & \text{QP}_1^2 \\ 1 & \text{QP}_2 & \text{QP}_2^2 \\ \vdots & \vdots & \vdots \\ 1 & \text{QP}_n & \text{QP}_n^2 \end{bmatrix},$$

$$\beta = \begin{bmatrix} \beta_{i,c,1,0} \\ \beta_{i,c,1,1} \\ \beta_{i,c,1,2} \end{bmatrix},$$

and $\varepsilon$ denotes an $n \times 1$ independent, identically distributed random noise vector. From equation (3), we obtain least-squares estimates using:

$$\beta = (X^T X)^{-1} X^T y$$

where we note that $(X^T X)^{-1} X^T$ only depends on QP.

To assess the accuracy of any given model, we compute the adjusted $R^2$ for each model. To define the adjusted $R^2$, we first need to define the residual sum of squares error (RSS) using:

$$\text{RSS} = \sum_{m=1}^{m=n} \left( \log(\text{VMAF}_m) - \beta_{i,c,1,0} - \beta_{i,c,1,1} \cdot \text{QP}_m - \beta_{i,c,1,2} \cdot \text{QP}_m^2 \right)^2.$$

We then compute the total sum of squares (TSS) using:

$$\text{TSS} = \sum_{m=1}^{m=n} (\log(\text{VMAF}_m) - \bar{y})^2$$

where the average output is given by:

$$\bar{y} = \frac{1}{n} \sum_{m=1}^{m=n} \log(\text{VMAF}_m).$$

We are now ready to define the adjusted $R^2$ using

$$\text{Adjusted } R^2 = 1 - \frac{\text{RSS}/(n-d-1)}{\text{TSS}/(n-1)} \quad (4)$$

where $n$ denotes the number of QP values and $d = 3$, which represents the number of non-zero $\beta$ values used in our model. For a perfect model fit, RSS=0, and we have that the adjusted $R^2$ is 1. It is also important to note that the definition of the adjusted $R^2$ takes into account the degree of the polynomial fit. For higher-order polynomials, the same value of RSS will produce a lower value for the adjusted $R^2$. Hence, by maximizing the adjusted $R^2$, we can also determine if higher-order polynomials are justified. In general, stepwise logistic regression refers to the process of selecting the polynomial order that maximizes the adjusted $R^2$ that balances the use of higher degree polynomials versus the degree of fit.

Inverting the regression model is easy to do. For example, for any given VMAF value, we use

$$\log(VMAF) = \beta_{i,c,1,1} - \beta_{i,c,1,2} \cdot \text{QP} - \beta_{i,c,1,3} \cdot \text{QP}^2$$

to determine the QP value that would deliver this video quality. In general, the determined QP value would not be an integer. Hence, we will need to round up the real-valued QP value to determine an integer solution.

*Multi-objective Optimization*

An exhaustive evaluation of all QP values for all possible configurations is not needed. Instead, we only need to consider values that are optimal in the multi-objective sense. In this section, we explain how this is accomplished.

Formally, we write that we are interested in solving the multi-objective optimization problem expressed as:

$$\max_{c,QP} (\text{VMAF}, \text{PSNR}, -B, \text{FPS}). \quad (5)$$

To solve equation (4), let the collection of points generated by all possible encodings be expressed as:

$$(\text{VMAF}_m, \text{PSNR}_m, B_m, FPS_m) \text{ for } m = 1, 2, \ldots, N.$$

Then, we eliminate an encoding $m = p$ if there is at least one other encoding for $m = k$ ($k \neq p$) that is better in all objectives as given by:

$$(\text{VMAF}_k \geq \text{VMAF}_p) \text{ and } (\text{PSNR}_k \geq \text{PSNR}_p) \text{ and }$$
$$(B_k \leq B_p) \text{ and } (\text{FPS}_k \geq \text{FPS}_p). \quad (6)$$

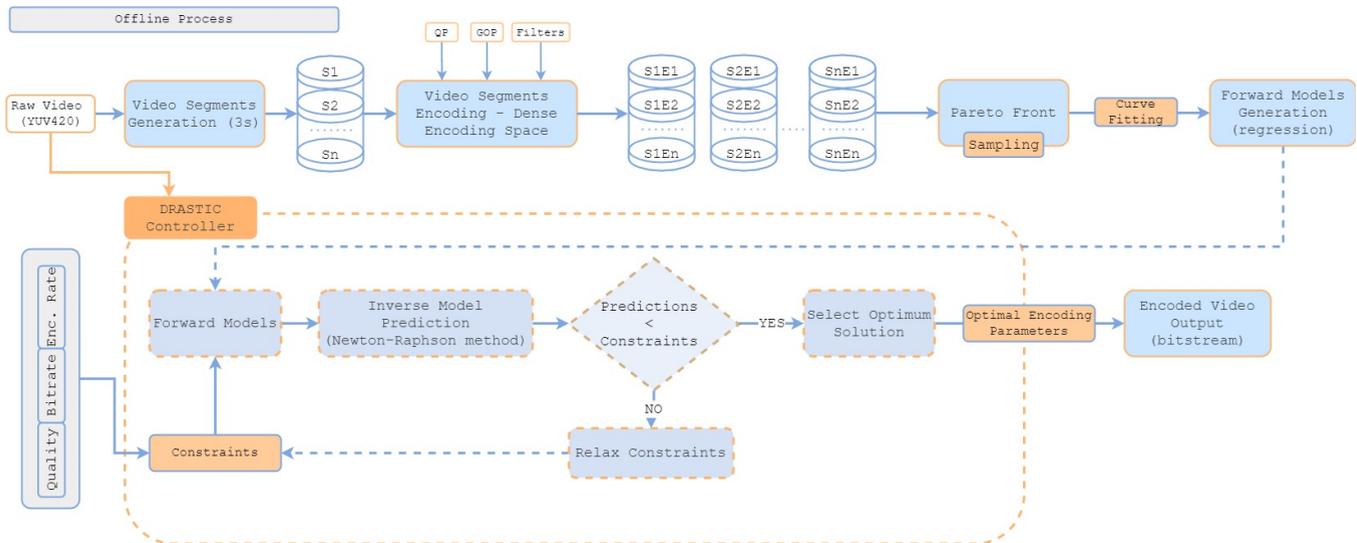

**FIGURE 2. Adaptive video encoding abstract system architecture.**
Raw video files are segmented into chucks of 3 seconds in duration to capture differences in content and characteristics within the same video sequence. A dense encoding space is then used to produce a sufficient sample of video instances per examined video codec. The approach relies on the Pareto front to apply curve fitting using regression, in order to generate forward prediction models per optimization objectives, namely video quality, bitrate, and encoding rate. During real-time video streaming, given a set of constrains and mode of operation, the Newton method is employed to predict an encoding configuration that satisfies these constraints. In case the encoding parameters prediction set is empty, the constraints are relaxed until a matching configuration is provided.

To generate the Pareto front, we compare all points against each other using a double for-loop through and eliminate points based on equation (6). We use the term Pareto front to describe the remaining encodings that do not get eliminated through the process.

### Formulating and Solving Constrained Optimization Problems

We summarize the constraint optimization modes that can be solved using our derived models. We note that all of the solutions lie on the Pareto front. To see this, note that if they did not lie on the Pareto front, we would then be able to select another encoding that is better in every respect.

We begin by establishing notation (also see Table III). Let VQ denote the video quality mode metric (e.g., VMAF). We would like to impose a minimum video quality bound as given by $VQ \geq VQ_{min}$. Similarly, for maximizing performance, we require that FPS, the achieved frames per second, must be above a minimum rate of performance as given by $FPS \geq FPS_{min}$. Also, we want to bound the encoded bitrate based on the available bitrate using: $B \leq B_{max}$. In what follows, we can define the constraint optimization problems based on this notation.

We next formulate and solve the constrained optimization problems that will provide optimal video encoding delivery. We formulate the *minimum bitrate mode* using:
$$\min B \quad (7)$$
subject to $(VQ \geq VQ_{min})$ and $(FPS \geq FPS_{min})$.
To solve equation (7), we need to consider encoding configurations that satisfy the constraints $(VQ \geq VQ_{min})$ and $(FPS \geq FPS_{min})$. Among these encodings, we then select the one that gives the minimum bitrate.

Similarly, we define the *maximum video quality* mode using:
$$\max VQ \quad (8)$$
subject to $(B \leq B_{max})$ and $(FPS \geq FPS_{min})$.
Then, the *maximum encoding performance* mode is defined using:
$$\max FPS \quad (9)$$
subject to $(B \leq B_{max})$ and $(VQ \geq VQ_{min})$.

We solve (8) and (9) by maximizing VQ or FPS over the encodings that meet our constraints.

#### 2) ENCODING CONFIGURATIONS FOR FORWARDING PREDICTION MODELS GENERATION

The encoding configurations for the forward prediction models are summarized in Tables IV.A-IV.C. More specifically, for each of the examined codecs, namely x265, VP9, and SVT-AV1, a total of 200, 200, and 252 video instances are generated, respectively, per investigated video segment. A similar configuration setup is used for all three encoders, which is tuned towards real-time performance. Different encoding structures and the use of deblocking filters to enhance quality are further considered. The objective here is to demonstrate the universality of the proposed codec agnostic adaptation framework.

#### 3) FORWARD PREDICTION MODELS USING LINEAR REGRESSION

As we described earlier, we compute forward prediction models to determine the mapping from the encoding parameters to the objectives of video quality, bitrate, and encoding time, in a live video streaming session, without having to actually encode the video. To succeed in this task, the available knowledge emanating from the above-described offline process used to generate a dense encoding space is required.

TABLE IV.A
X265 CONFIGURATIONS FOR ULTRAFAST PRESET

| Parameter | Value | Parameter | Value |
|---|---|---|---|
| Profile | Ultrafast | SAO filter | On/Off |
| Encoding Structure | B2, B3, B4, B6, ZL | DBF filter | On/Off |
| GOP structure | Open/Close | Tune | PSNR |
| QP | 18-45[a] | CTU | 64 |
| **Total configurations per video segment** | | | **200** |

[a]Increments of 3; Key-Interval = <frame-rate>;
ZL = Zero latency (IPPP); B2/B3/B4/B6 = number of consecutive B frames between P frames; SAO = Sample Adaptive Offset; DBF = Deblocking Filter; CTU = Coding Tree Unit [17];

TABLE IV.B
VP9 CONFIGURATIONS FOR REALTIME PRESET

| Parameter | Value | Parameter | Value |
|---|---|---|---|
| Preset | Realtime | arnr-max Frames | 7 |
| Encoding Structure | ALT0, ALT1, ALT2, ALT4, ALT6 | arnr-strength | 5 |
| QP | 16-52[a] | arnr-type | 3 |
| DBF filter | On/Off | Tune | PSNR |
| **Total configurations per video segment** | | | **200** |

[a]Increments of 4; Key-Interval = <frame-rate>;
ALT0/1/2/4/6: number of alternate reference frames [18];

TABLE IV.C
SVT-AV1 CONFIGURATIONS FOR ENCODE MODE 7

| Parameter | Value | Parameter | Value |
|---|---|---|---|
| Encode Mode | 7 | arnr-max Frames | 7 |
| Encoding Structure | HL3ALT0, HL3ALT2, HL3ALT8, HL4ALT0, HL4ALT2, HL4ALT8 | arnr-strength | 5 |
| QP | 16-52[a] | arnr-type | 3 |
| DBF filter | On/Off | Tune | PSNR |
| **Total configurations per video** | | | **252** |

[a]Increments of 4; Key-Interval = <frame-rate>;
HL3/4: Hierarchy level (number of temporal layers); ALT0/1/2/4: number of alternate reference frames [16];

Linear and logistic regression up to third order polynomials were used to determine the most suitable models as functions of the encoding configurations given in Table IV per examined video codec. Furthermore, stepwise regression was employed to optimize the trade-off between the cross-validated residual error and model complexity, hence limiting overfitting with more complex models.

### 4) REAL-TIME ENCODING USING MULTI-OBJECTIVE OPTIMIZATION

Adaptive video encoding for real-time video delivery applications leverages the codec-agnostic algorithm described in Fig. 3 that implements the abstract system architecture highlighted in Fig. 2. The basic algorithm is broken into two parts. First, the forward prediction models are computed offline for each video segment. Second, the forward prediction models are used to adaptively encode each video segment for a pre-processed video.

For computing the forward prediction models, we apply stepwise regression for modeling the Pareto front of the VMAF, PSNR, bitrate, and encoding rate of each video

```
Algorithm 1: Compute Forward Regression Models
  input : Video database, Video Codecs, and Enc. Parameters.
  output: Forward regression models for each video segment
1 for each video in the video database do
2   for each video segment, encoding configurations do
3     Encode the video segment using encoding configuration
           Compute Pareto front for VMAF, PSNR, FPS, Bitrate
         if this is the second or later video segment then
4          | Compute error for using previous regression model
5        end
6        if this is the first video segment or the error is high then
7          Compute forward regression models
             Store forward regression models for each video segment
8        else
9          Reuse regression models from previous video segment
10       end
11   end
12 end
```

```
Algorithm 2: Adaptive Video Encoding
  input : Video, Forward regression models, optimization modes
  output: Optimally Encoded Video Segments
1 while more video GOP segments to encode do
2   Apply regression models to compute
          Optimal Encoding Parameters C_Opt and QP_Opt
          for the required optimization mode and constraints
3   if either QP_Opt or C_Opt is out of range then
4     Update feasible constraints and find
             new estimates of C_Opt and QP_Opt
         Constrain QP_Opt to be within +/-4 of neighboring QP
         Enforce QP and C_Opt within valid ranges
5   end
6   Encode the video using C_Opt and QP_Opt
         Compute PSNR_Opt, VMAF_Opt, FPS_Opt, Bitrate_Opt for current GOP segment
7 end
```

**FIGURE 3.** Segment-based adaptive video encoding.

segment of each video. Here, we note that the goal is to summarize all of the encoding options with simple regression models that are fast to process during real-time video delivery. Thus, instead of storing the encoded video segments, we store the parameters of the forward regression models.

After storing the forward regression models for the first video segment, we also consider the possibility of reusing the regression models from the previous video segment. Here, we reuse the forward regression models from the previous segment if they can accurately predict the encoding for the current video segment (see second if statement in Fig. 3). To reuse the models, we require that the fitted regression model error does not exceed the maximum values of 5% for video quality, 10% for bitrate, and 10% for encoding time for any one of the encodings. Hence, for real-time applications, our approach can reduce the overall computational complexity while sacrificing model accuracy.

Then, for real-time video delivery, each video is broken into the same video segments like the ones used for computing the forward models. We then retrieve the forward models for each segment and use them to determine points that satisfy the optimization modes. We use the procedure outlined in section II.C 1) to compute the QP_opt parameter for each chosen encoding configuration C_opt. Here, we note that the selected QP_opt parameter is continuously valued. Thus, we quantize QP_opt. In the event that the pair QP_opt and C_opt are not valid, we relax the constraints to get a valid solution. In this case, we allow QP to vary from -

TABLE V
BITRATE SAVINGS[1] PER VIDEO RESOLUTION AND CODEC COMPARISON PAIRS USING BD-PSNR

| VVC vs | Bitrate Savings per Video Resolution of VVC vs {AV1, x265, VP9} | | | | | | | |
|---|---|---|---|---|---|---|---|---|
| | 240p | 432p | 480p | 720p | 1080p | 1600p[2] | Median | Average ($\sigma^2$) |
| AV1 | 53.6% | 56.2% | 56% | 44.5% | 49.8% | 50.2% | 51.90% | 51.7% (4.5%) |
| x265 | 71.1% | 67.5% | 70.3% | 57.8% | 67% | 65.3% | 67.3% | 66.6% (4.8%) |
| VP9 | 71.8% | 75% | 73% | 74.7% | 75.8% | 69.2% | 73.9% | 73.3% (2.5%) |
| AV1 vs | Bitrate Savings per Video Resolution of AV1 vs {x265, VP9} | | | | | | | |
| | 240p | 432p | 480p | 720p | 1080p | 1600p | Median | Average ($\sigma^2$) |
| x265 | 38.8% | 23.3% | 31.1% | 35.8% | 32.6% | 30.3% | 31.8% | 32% (5.3%) |
| VP9 | 40.9% | 36.5% | 38.4% | 62.3% | 51% | 38.7% | 39.8% | 44% (10%) |
| X265 vs | Bitrate Savings per Video Resolution of x265 vs VP9 | | | | | | | |
| | 240p | 432p | 480p | 720p | 1080p | 1600p | Median | Average ($\sigma^2$) |
| VP9 | 3.7% | 25.3% | 11.1% | 39.3% | 27.6% | 12.5% | 18.9% | 19.8% (13.1%) |

TABLE VI
BITRATE SAVINGS PER VIDEO RESOLUTION AND CODEC COMPARISON PAIRS USING BD-VMAF

| VVC vs | Bitrate Savings per Video Resolution of VVC vs {AV1, x265, VP9} | | | | | | | |
|---|---|---|---|---|---|---|---|---|
| | 240p | 432p | 480p | 720p | 1080p | 1600p | Median | Average ($\sigma^2$) |
| AV1 | 60% | 64.8% | 59.6% | 63% | 54.2% | 68.2% | 61.5% | 61.6% (4.8%) |
| x265 | 77.3% | 68.9% | 71.5% | 73.4% | 59.8% | 67.8% | 70.2% | 69.8% (6%) |
| VP9 | 75.1% | 74.9% | 79.8% | 76.4% | 67.8% | 68% | 75% | 73.8% (4.8%) |
| AV1 vs | Bitrate Savings per Video Resolution of AV1 vs {x265, VP9} | | | | | | | |
| | 240p | 432p | 480p | 720p | 1080p | 1600p | Median | Average ($\sigma^2$) |
| x265 | 44.23% | 7.3% | 27.06% | 29.85% | 13.73% | 29% | 28.03% | 25.60% (11.98%) |
| VP9 | 34.8% | 12.8% | 50.6% | 38.64% | 26.77% | 38% | 36.41% | 34.00% (11.7%) |
| X265 vs | Bitrate Savings per Video Resolution of x265 vs VP9 | | | | | | | |
| | 240p | 432p | 480p | 720p | 1080p | 1600p | Median | Average ($\sigma^2$) |
| VP9 | -18.1%[3] | 17.9% | 32.5% | 16.2% | 17.8% | 16% | 18% | 19% (5.8%) |

[1] Bitrate savings have been rounded up to one decimal
[2] 1600p average values are based on two videos.
[3] VP9 reduces bitrate demands by 18.1% compared to x265 {240p, BD-VMAF}

4 to +4 from QP_opt and consider alternative encodings until we identify valid encoding parameters.

## III. RESULTS

In what follows, we compare the compression effectiveness of different video codecs using both objective and subjective video quality assessment. We present and discuss results using PSNR and VMAF for a variety of video resolutions. Then, we describe adaptive video encoding results to demonstrate the advantages of our adaptive approach over non-adaptive approaches.

We provide a summary of our software and hardware platforms to support the reproducibility of our results. We implemented our methods on a Windows 10 Dell Precision Tower 7910 Server 64-bit platform with Intel(R) Xeon(R) Processor E5-2630 v3 (8 cores, 2.4GHz). In terms of software, we used SVT-AV1 version 0.7, VP9 version 1.8, and x265 version 2.0.

### A. COMPARATIVE PERFORMANCE EVALUATION OF VIDEO COMPRESSION STANDARDS

#### 1) OBJECTIVE VIDEO QUALITY ASSESSMENT

We provide comprehensive results comparing all video codecs against each other using BD-PSNR in Table V and using BD-VMAF in Table VI. For better visualization, the results are also plotted in Fig. 4 for both BD-PSNR and BD-VMAF.

Clearly, VVC achieves the best compression efficiency compared to all rival standards, followed by SVT-AV1. X265 enhances compression efficiency compared to VP9 in all but the lowest examined video resolution, namely 240p, when VMAF is used to compute the BD-rate gains.

More specifically, VVC reduces bitrate demands compared to SVT-AV1 by ~51.7% on average based on PSNR scores, climbing to ~61.6% when VMAF ratings are used. There is approximately a 10% difference in favor of VVC when VMAF is employed, which calls for further validation in conjunction with large-scale subjective evaluation studies (see also subjective VQA, below). Significant bitrate gains are observed when VVC is compared against x265 and VP9. With respect to x265, average bitrate gains are comparable and extend to ~66.6% and ~69.9% for PSNR and VMAF, respectively. When compared against VP9, results from both algorithms are also in agreement, documenting reduced bitrate requirements between ~73% and ~74%.

SVT-AV1 achieves significant compression performance gains over x265, as depicted in Tables V and VI. Bitrate gains using BD-PSNR are in the order of ~32% and slightly reduced to 25.6% when BD-VMAF is used. Likewise,

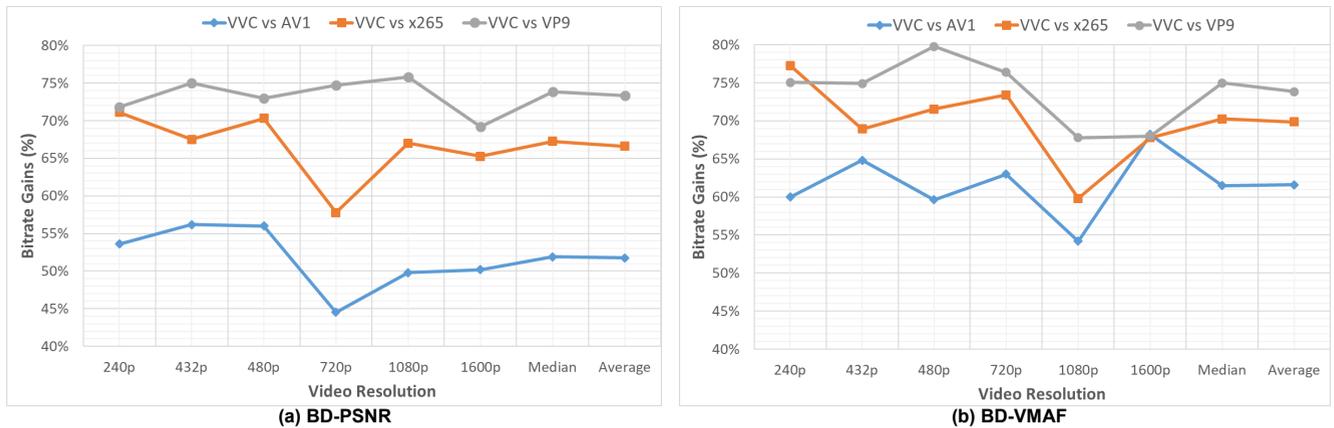

FIGURE 4. Average bitrate demands reductions of VVC against {AV1, x265, VP9} across the video resolution ladder using (a) BD-PSNR and (b) BD-VMAF.

TABLE VII
SUBJECTIVE AND OBJECTIVE VQA CORRELATION PER VIDEO CODEC FOR THE MIXED BAG DATASET.

| Correlation Metrics per objective VQA method | Video Codec and Objective VQA Metrics | | | | | |
|---|---|---|---|---|---|---|
| | SVT-AV1 | | VP9 | | x265 | |
| | PSNR | VMAF | PSNR | VMAF | PSNR | VMAF |
| SROCC | 0.64 | **0.75** | 0.63 | **0.74** | 0.61 | **0.635** |
| POCC | 0.623 | **0.78** | 0.6 | **0.70** | 0.58 | **0.633** |

SROCC: Spearman Rank Order Correlation Coefficient; POCC: Pearson Correlation Coefficient;

increased performance over its predecessor, VP9, is documented, reaching ~44% and 34%, for PSNR and VMAF computed BD-rate gains, respectively. Again, there is a noticeable difference between the two objective scores. As discussed in the subjective evaluation sub-section below, VMAF's high correlation to perceptual quality ratings for both AV1 and VP9 suggests that the bitrate gains using VMAF are, in fact, more realistic.

In the last set of codec comparison pairs, x265 supersedes VP9 in compression efficiency, recording average gains between 19%-20% for both VQA algorithms used. However, it is important to highlight that results show a great variation across the video resolution ladder. Especially when PSNR is used, the standard deviation measured is ~13%. Moreover, at the lower 240p end, VP9 is found to be more efficient in terms of the VMAF scores.

2) SUBJECTIVE VIDEO QUALITY ASSESSMENT

We validate our results using subjective video quality assessment. In previously published studies, the HEVC/H.265 standard matched the targeted bitrate gains of ~50% over its predecessor, H.264/AVC, based on subjective ratings, as opposed to objective assessment, which actually documented lesser gains [31]. For our purposes, we note that it is impractical to conduct subjective video quality studies on very large datasets. Instead, we conducted our study on a representative sample of videos that capture different video content, compression levels, and video characteristics (i.e., resolution and framerate). This sample is then evaluated and used to compute the correlation between subjective and objective ratings.

The subjective VQA results' correlation to the objective scores is summarized in Table VII. The subset of the investigated datasets totaled 108 video instances that were assessed by 32 human subjects in three different sessions. Each session corresponded to videos encoded with SVT-AV1, x265, and VP9 compression standards. Unfortunately, assessing VVC videos was not possible at the time that the evaluation sessions took place due to the unavailability of a reliable VVC player. As a result, we only report the correlation between subjective and objective scores of codecs belonging to the aforementioned groups.

Table VII demonstrates the correlation of the subjective evaluation scores to both PSNR and VMAF objective VQA metrics per investigated video codec. Two widely used correlation indices were employed for this purpose, namely the Spearman rank-order correlation coefficient (SROCC) and the Pearson linear correlation coefficient (PLCC). The VMAF algorithm achieved a significantly higher correlation to the subjective ratings than PSNR for all three examined codecs and both correlation coefficients. The latter observation strongly suggests that documented bitrate gains using BD-VMAF highlighted in Table VI, are in fact, more trustworthy and better reflect the performance comparison of the examined video codecs. Moreover, VMAF's correlation results justify its wide adoption over the past few years amongst the research community and the industry. On the other hand, PSNR failed to adequately capture the user-perceived quality across all codecs and correlation indices. In particular, PLCC recorded values between 0.58 and 0.623, while SROCC slightly better correlation ratings from

TABLE VIII
FORWARD PREDICTION MODELS ADJUSTED R SQUARE PER OPTIMIZATION OBJECTIVE – x265 CODEC.

| Objectives | BasketballDrive - 1920x1080@50fps | | | | | | PartyScene – 832x480@50fps | | | | | |
|---|---|---|---|---|---|---|---|---|---|---|---|---|
| | Linear | | | Quadratic | | | Linear | | | Quadratic | | |
| | Min | Med | Max | Min | Med | Max | Min | Med | Max | Min | Med | Max |
| *PSNR* | 0.987 | 0.992 | 0.995 | 0.989 | 0.997 | 0.998 | 0.996 | 0.999 | 1 | 0.996 | 0.999 | 1 |
| *Bitrate* | 0.983 | 0.990 | 0.994 | 0.997 | 0.998 | 0.999 | 0.981 | 0.991 | 0.996 | 0.999 | 1 | 1 |
| *FPS* | 0.930 | 0.951 | 0.972 | 0.956 | 0.987 | 0.996 | 0.675 | 0.946 | 0.977 | 0.772 | 0.946 | 0.975 |
| *VMAF* | 0.765 | 0.799 | 0.850 | 0.993 | 0.996 | 0.998 | 0.762 | 0.784 | 0.821 | 0.973 | 0.984 | 0.99 |

| Objectives | BlowingBubbles – 480x240@50fps | | | | | | SunFlower - 768x432@25fps | | | | | |
|---|---|---|---|---|---|---|---|---|---|---|---|---|
| | Linear | | | Quadratic | | | Linear | | | Quadratic | | |
| | 0.995 | 0.999 | 1 | 0.998 | 0.999 | 1 | 0.998 | 0.999 | 1 | 0.998 | 0.999 | 1 |
| *PSNR* | 0.985 | 0.995 | 0.999 | 0.999 | 1 | 1 | 0.995 | 1 | 1 | 0.998 | 1 | 1 |
| *Bitrate* | 0.861 | 0.922 | 0.951 | 0.893 | 0.937 | 0.963 | 0.860 | 0.932 | 0.984 | 0.846 | 0.928 | 0.990 |
| *FPS* | 0.741 | 0.811 | 0.831 | 0.981 | 0.988 | 0.992 | 0.663 | 0.722 | 0.763 | 0.968 | 0.978 | 0.988 |
| *VMAF* | 0.995 | 0.999 | 1 | 0.998 | 0.999 | 1 | 0.998 | 0.999 | 1 | 0.996 | 0.999 | 1 |

FORWARD PREDICTION MODELS ADJUSTED R SQUARE PER OPTIMIZATION OBJECTIVE – AV1 CODEC.

| Objectives | BQTerrace - 1920x1080@60fps | | | | | | Kristin and Sara - 1280x1080@50fps | | | | | |
|---|---|---|---|---|---|---|---|---|---|---|---|---|
| | Linear | | | Quadratic | | | Linear | | | Quadratic | | |
| | Min | Med | Max | Min | Med | Max | Min | Med | Max | Min | Med | Max |
| *PSNR* | 0.928 | 0.953 | 0.971 | 0.99 | 0.997 | 0.999 | 0.972 | 0.978 | 0.989 | 0.996 | 0.997 | 0.998 |
| *Bitrate* | 0.992 | 0.996 | 0.998 | 0.994 | 0.998 | 0.999 | 0.991 | 0.995 | 0.998 | 0.998 | 0.999 | 0.999 |
| *FPS* | 0.917 | 0.968 | 0.989 | 0.975 | 0.987 | 0.996 | 0.712 | 0.776 | 0.886 | 0.887 | 0.952 | 0.967 |
| *VMAF* | 0.694 | 0.791 | 0.976 | 0.945 | 0.972 | 0.999 | 0.865 | 0.888 | 0.900 | 0.991 | 0.993 | 0.994 |

| Objectives | Tango – 1280x720@30fps | | | | | | Pedestrian – 768x432@25fps | | | | | |
|---|---|---|---|---|---|---|---|---|---|---|---|---|
| | Linear | | | Quadratic | | | Linear | | | Quadratic | | |
| *PSNR* | 0.964 | 0.968 | 0.969 | 0.999 | 0.999 | 0.999 | 0.980 | 0.983 | 0.985 | 0.986 | 0.989 | 0.991 |
| *Bitrate* | 0.997 | 0.999 | 0.999 | 0.999 | 0.999 | 0.999 | 0.995 | 0.996 | 0.997 | 0.999 | 0.999 | 0.999 |
| *FPS* | 0.955 | 0.963 | 0.970 | 0.955 | 0.963 | 0.970 | 0.956 | 0.970 | 0.978 | 0.957 | 0.970 | 0.978 |
| *VMAF* | 0.765 | 0.793 | 0.816 | 0.993 | 0.994 | 0.996 | 0.747 | 0.781 | 0.808 | 0.988 | 0.991 | 0.994 |

FORWARD PREDICTION MODELS ADJUSTED R SQUARE PER OPTIMIZATION OBJECTIVE – VP9 CODEC.

| Objectives | BasketballDrill - 1920x1080@50fps | | | | | | BlowingBubbles - 480x240@50fps | | | | | |
|---|---|---|---|---|---|---|---|---|---|---|---|---|
| | Linear | | | Quadratic | | | Linear | | | Quadratic | | |
| *PSNR* | 0.813 | 0.990 | 0.991 | 0.816 | 0.998 | 0.999 | 0.944 | 0.993 | 0.994 | 0.943 | 0.998 | 0.998 |
| *Bitrate* | 0.829 | 0.991 | 0.992 | 0.830 | 0.993 | 0.993 | 0.925 | 0.987 | 0.989 | 0.927 | 0.996 | 0.996 |
| *FPS* | 0.806 | 0.902 | 0.952 | 0.820 | 0.902 | 0.968 | 0.762 | 0.882 | 0.946 | 0.766 | 0.876 | 0.949 |
| *VMAF* | 0.524 | 0.707 | 0.737 | 0.780 | 0.984 | 0.986 | 0.621 | 0.744 | 0.798 | 0.815 | 0.971 | 0.986 |

| Objectives | Kristin and Sara - 1280x1080@50fps | | | | | | SunFlower - 768x432@25fps | | | | | |
|---|---|---|---|---|---|---|---|---|---|---|---|---|
| | Linear | | | Quadratic | | | Linear | | | Quadratic | | |
| *PSNR* | 0.879 | 0.971 | 0.979 | 0.900 | 0.998 | 0.999 | 0.69 | 0.989 | 0.992 | 0.834 | 0.998 | 0.998 |
| *Bitrate* | 0.900 | 0.989 | 0.990 | 0.896 | 0.995 | 0.995 | 0.871 | 0.993 | 0.994 | 0.868 | 0.994 | 0.995 |
| *FPS* | 0.851 | 0.895 | 0.924 | 0.861 | 0.963 | 0.986 | 0.755 | 0.877 | 0.928 | 0.788 | 0.887 | 0.969 |
| *VMAF* | 0.800 | 0.848 | 0.869 | 0.926 | 0.986 | 0.988 | 0.477 | 0.725 | 0.835 | 0.655 | 0.965 | 0.986 |

<sup>a</sup>The adjusted $R^2$ values have been rounded to three significant digits. As a result, some values are rounded up to 1.

0.61 to 0.64, for the three video compression standards, investigated.

We obtained the highest correlations for the SVT-AV1 codec. More specifically, AV1 achieved a 0.78 and a 0.75 PLCC and SROCC correlation, respectively, to the VMAF scores. Interestingly, VP9 performed significantly better than the x265 codec, reaching a 0.74 correlation compared to 0.635 for the SROCC and 0.7 against 0.633 for the PLCC, both for VMAF scores. Given the gap from 0.6-0.7 to the ideal correlation of 1.0, despite the great progress achieved by VMAF, it is clear that there is a need for continued research on developing reliable VQA that better correlate to human interpretation. In that sequence, such metrics should achieve correlation to user ratings over 0.95 before they can be used as the sole criterion to measure user's perceived QoE. Clearly, larger studies are also needed to support such efforts, securing the objectivity of such algorithms across video codecs, video content, and video characteristics.

### B. ADAPTIVE VIDEO ENCODING VALIDATION

A representative subset of the proposed adaptive video encoding framework validation space is presented in the current section. We start by providing the computed forward prediction model equations. Then, we demonstrate specific use-case scenarios involving the x265 and SVT-AV1 video codecs and different video sequences. For each codec, the precise forward models' equations per examined video and selected mode of operation are presented, followed by the advantages of the proposed methods.

### 1) FORWARD PREDICTION MODELS EQUATIONS

In this section, we provide a summary of our regression models. We consider a total of 5 forward prediction models per objective for x265, 5 for VP9, and 6 for SVT-AV1.

*I. Forward Prediction Models Validation*

To demonstrate the efficiency of the generated forward prediction models, we provide results documenting the adjusted $R^2$ of the fitted models. As detailed in the methodology, a perfect fit results in an adjusted $R^2$ value of 1. Hence, the closer the adjusted $R^2$ is to 1, the higher is the model's accuracy. Table VIII tabulates the results for linear and quadratic logistic models by video, video codec, and optimization objective. The minimum, maximum, and median adjusted $R^2$ values are given, abstracted from the entire set of considered models per segment, and employed video coding structure (3 segments x 5 models for x265, VP9, and 6 for AV1). Only encoding structures with the best-performing filters option were considered, as described above. In this fashion, we can demonstrate how well the models fit or not for a certain video before used for adaptive, real-time video streaming purposes.

Due to space constraints, four selected videos per video codec, for a total of 12 videos, are summarized in Table VIII. The key message here is that the proposed methodology achieves robust models' fitting that can hence be confidently used for adaptive video encoding using the approach described by the pseudocode algorithm of Fig. 3. In particular, the majority of median adjusted $R^2$ values of linear models are significantly higher than 0.9, indicating strong fits for all objectives except for VMAF. For VMAF, quadratic models provide the best fits, which are used instead. The same holds for certain videos with respect to the FPS objective, such as the Kristin and Sara video AV1 models and the lower resolution VP9 models of BlowingBubbles and SunFlower videos. In all cases, however, adjusted $R^2$ median values, even for linear models, are higher than 0.7. Clearly, based on the depicted results, the proposed methods can be used to derive robust models over video datasets with diverse characteristics and content.

*II. Comparison of Linear and Quadratic Models*

Fig. 5 displays boxplots of adjusted $R^2$ median values of linear and quadratic models, per investigated video codec and optimization objective. Median values are drawn from the predicted forward models on a segment basis for each one of the four videos depicted in Table VIII. In other words, the median values are grouped only based on the encoding structure, contrary to Table VIII, where grouping per both segment and encoding structure was performed. Results reiterate the robust fit of the generated models that are able to capture a video's unique characteristics.

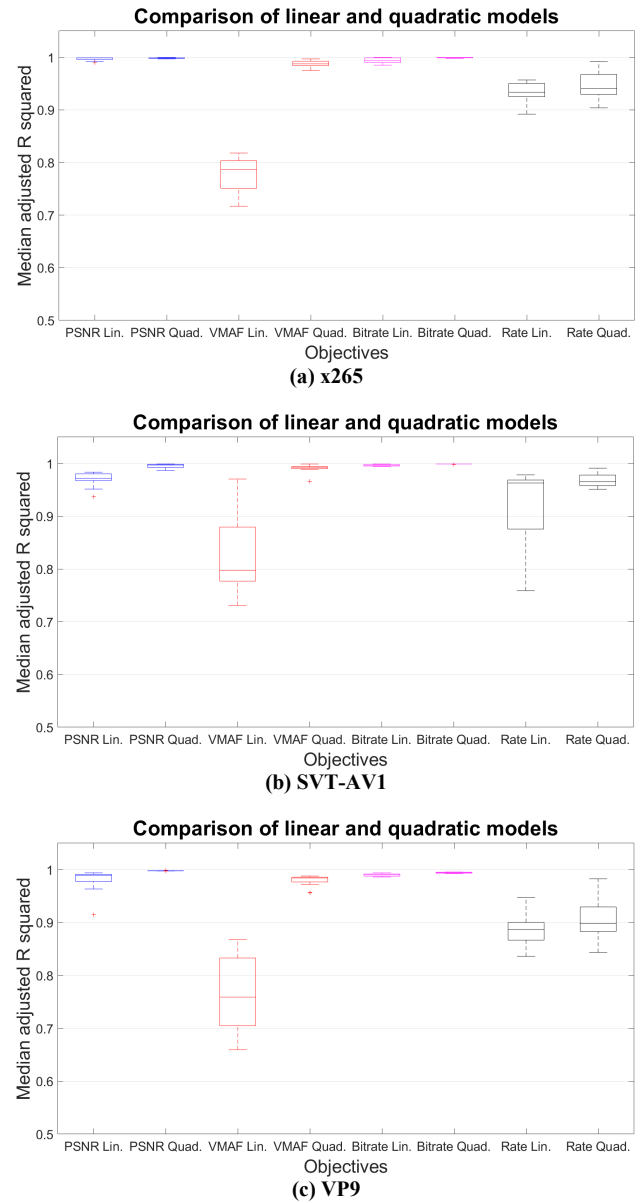

**FIGURE 5.** Boxplots depicting adjusted $R^2$ median values of linear and quadratic models, per investigated video codec and optimization objective. (a) x265, (b) SVT-AV1, and (c) VP9 video codecs.

As evident, for PSNR and Bitrate objectives, adjusted $R^2$ median values are virtually indistinguishable, hence favoring the use of linear models during adaptive video encoding. A strong fit is also depicted for the encoding rate (i.e., FPS) objective, despite the slightly lower adjusted $R^2$ median values and higher variation, especially for SVT-AV1 models. On the other hand, quadratic models are needed for the VMAF objective.

### 2) ADAPTIVE VIDEO ENCODING

We present two examples to demonstrate adaptive video coding. We selected the Cactus 1080p video of the HEVC dataset for demonstrating adaptive encoding using x265. For SVT-AV1, we selected the Pedestrian 432p video. For both videos, we present results for the minimum bitrate

TABLE IX
FORWARD PREDICTION MODELS COEFFICIENTS PER SEGMENT AND OPTIMIZATION OBJECTIVE – x265 FOR CACTUS 1080P VIDEO.

| | Segments | $c = B2$ | | | | $c = B3$ | | | | $c = B4$ | | | |
|---|---|---|---|---|---|---|---|---|---|---|---|---|---|
| | | $\beta_{i,c,j,0}$ | $\beta_{i,c,j,1}$ | $\beta_{i,c,j,2}$ | $R^2$ | $\beta_{i,c,j,0}$ | $\beta_{i,c,j,1}$ | $\beta_{i,c,j,2}$ | $R^2$ | $\beta_{i,c,j,0}$ | $\beta_{i,c,j,1}$ | $\beta_{i,c,j,2}$ | $R^2$ |
| $\log(PSNR_i)$ | Seg 0 | 3.89 | 0.0085 | 0.000044 | 0.990 | 3.86 | -0.0066 | -0.000074 | 0.990 | 3.86 | -0.0066 | -0.000073 | 0.990 |
| | Seg 1 | 3.88 | -0.0073 | -0.000100 | 0.996 | 3.86 | -0.0057 | -0.000100 | 0.996 | 3.85 | -0.0054 | -0.000100 | 0.996 |
| | Seg 2 | 3.89 | -0.0079 | -0.000100 | 0.997 | 3.87 | -0.0063 | -0.000100 | 0.996 | 3.86 | -0.0061 | -0.000100 | 0.996 |
| $\log(VMAF_i)$ | Seg 0 | 3.84 | 0.0670 | -0.001499 | 0.980 | 3.80 | 0.0696 | -0.001548 | 0.990 | 3.82 | 0.0685 | -0.001532 | 0.990 |
| | Seg 1 | 3.86 | 0.0664 | -0.001500 | 0.989 | 3.84 | 0.0679 | -0.001500 | 0.992 | 3.83 | 0.0681 | -0.001500 | 0.992 |
| | Seg 2 | 3.81 | 0.0698 | -0.001600 | 0.987 | 3.81 | 0.0701 | -0.001600 | 0.991 | 3.81 | 0.0704 | -0.001600 | 0.992 |
| $\log(Bitrate_i)$ | Seg 0 | 16.97 | -0.3374 | 0.002444 | 0.990 | 16.65 | -0.3198 | 0.002179 | 0.990 | 16.52 | -0.3131 | 0.002101 | 0.990 |
| | Seg 1 | 17.16 | -0.3530 | 0.002700 | 0.998 | 16.91 | -0.3405 | 0.002000 | 0.998 | 16.75 | -0.3318 | 0.002400 | 0.998 |
| | Seg 2 | 16.89 | -0.3298 | 0.002300 | 0.998 | 16.65 | -0.3184 | 0.0022000 | 0.998 | 16.43 | -0.3054 | 0.002000 | 0.998 |
| $\log(FPS_i)$ | Seg 0 | 0.69 | 0.1581 | -0.001727 | 0.970 | 0.71 | 0.1530 | -0.001585 | 0.960 | 0.40 | 0.1740 | -0.001899 | 0.930 |
| | Seg 1 | 0.56 | 0.1811 | -0.002200 | 0.980 | 0.77 | 0.1593 | -0.001700 | 0.967 | 1.02 | 0.1410 | -0.001500 | 0.957 |
| | Seg 2 | 0.80 | 0.1615 | -0.001800 | 0.990 | 1.02 | 0.1418 | -0.001400 | 0.969 | 1.16 | 0.1334 | -0.001300 | 0.961 |

$j \in \{\text{PSNR, VMAF, BITRATE, FPS}\}$

The $\beta_{i,c,j,0}$, $\beta_{i,c,j,1}$, and $\beta_{i,c,j,2}$ coefficients have been rounded up to 2, 4, and 6 significant digits, respectively.
$R^2$ refers to the Adjusted $R^2$. The adjusted $R^2$ values have been rounded to three significant digits.

optimization mode, extending our results for bitrate savings across different codecs.

For both examples, we break each video into three-second segments. Then, we compute forward prediction models per segment as depicted in Tables IX and XIII. In addition to the model coefficients, we also report the adjusted R² value for each video segment. Based on the high values for the adjusted R² values, we can see that the forward regression models provide excellent fits for all video segments. As described in Figs. 2 and 3, we consider reusing the forward models from the first video segment provided the model errors remain below some maximum values. In the present examples, the high proximity of the forward regression models coefficients between segments linked with the high adjusted R² values justifies the selection of the first segment's models for use throughout the video's remaining segments. Hence, for both scenarios described next, the forward regression models from the videos' first segment are used for subsequent segments. Here, it also worth noticing the very low values of the $\beta_{i,c,j,2}$ coefficient, which essentially translates to the use of linear models over quadratic models, for the specific examples (see also Table VIII and Fig. 5).

### I. Minimum Bitrate Demands using x265

To compare against static approaches, we consider encoding the Cactus 1080p video of the HEVC dataset based on YouTube recommendations described in [36]. We encode each video segment using constant quality encoding and x265 default encoding parameters. We only vary the selected QP parameter to approximate the bitrate recommendations as tabulated in Table X. We then average quality (PSNR and VMAF), bitrate, and FPS over the first three segments composing the Cactus video sequence and use these values as constraints to our multi-objective optimization approach described next. We choose to only use the first three segments to facilitate simpler results reporting, given that the Cactus video sequence duration is 10 seconds, and each segment corresponds to 3 seconds. Note that the proposed models can be trained over any segment duration.

Table XI demonstrates the benefits of using the proposed adaptive video encoding approach. For the minimum bitrate optimization mode depicted, the objective is to minimize bitrate subject to the quality and encoding rate constraints as given in equation (3). We use PSNR as the quality constraint and require 37.15 dB that matches the average value

TABLE X
STATIC MODE (DEFAULT) – YOUTUBE RECOMMENDED SETTINGS – x265

| Seg. ID | CQP | GOP | FIL | Bitrate (kbps) | FPS | PSNR (dB) | VMAF |
|---|---|---|---|---|---|---|---|
| Seg. 0 | 28 | B3 | ON | 11,130 | 57.27 | 37.12 | 92.48 |
| Seg. 1 | 28 | B3 | ON | 10,148 | 59.31 | 37.23 | 92.85 |
| Seg. 2 | 28 | B3 | ON | 11,314 | 58.73 | 37.09 | 92.42 |
| Avg | | | | 10,864 | 58.44 | 37.15 | 92.58 |

TABLE XI
x265 MINIMUM BITRATE OPTIMIZATION MODE - CACTUS 1080P VIDEO[1].

| Enc. Seg. | GOP | QP | Bitrate (kbps) | FPS | PSNR (dB) | VMAF |
|---|---|---|---|---|---|---|
| Seg. 0 | B4 | 29 | 9,323 | 49.21 | 37.07 | 90.58 |
|  |  | constraints |  | ≥50 | ≥37.15 |  |
| Seg. 1 | B4 | 28 | 10,179 | 47.69 | 37.54 | 92.26 |
|  |  | constraints |  | ≥50 | ≥37.15 |  |
| Seg. 2 | B3 | 29 | 9,468 | 51.09 | 37.12 | 90.81 |
|  |  | constraints |  | ≥50 | ≥37.15 |  |
| Avg |  |  | 9,656 | 49.33 | 37.24 | 91.22 |
| ΔGain |  |  | 11.1% | (-9.11) | 0.10 | (-1.37) |

[1]Predicted values using the models depicted in Table VII.

TABLE XII
x265 MINIMUM BITRATE OPTIMIZATION MODE - CACTUS 1080P VIDEO[1]. JUST NOTICEABLE DIFFERENCE 3-POINT VMAF EXAMPLE.

| Enc. Seg. | GOP | QP | Bitrate (kbps) | FPS | PSNR (dB) | VMAF |
|---|---|---|---|---|---|---|
| Seg. 0 | B4 | 30 | 7,894 | 56.73 | 36.75 | 89.22 |
|  |  | constraints |  | ≥50 |  | ≥89.58 |
| Seg. 1 | B4 | 30 | 7,210 | 58.12 | 36.88 | 89.67 |
|  |  | constraints |  | ≥50 |  | ≥89.58 |
| Seg. 2 | B3 | 31 | 6,788 | 61.03 | 36.38 | 87.67 |
|  |  | constraints |  | ≥50 |  | ≥89.58 |
| Avg |  |  | 7,297 | 58.63 | 36.67 | 88.85 |
| ΔGains |  |  | 32.8% | 0.19 | (-0.48) | (-3.73) |

[1]Predicted values using the models depicted in Table VII.

achieved by the YouTube recommendations are given in Table X. We note that quality constraint can be either PSNR or VMAF, and the appropriate model can be selected and invoked accordingly. However, we report output values for both quality metrics to facilitate consistency between experiments. In terms of encoding rate, the constraint is set to match the real-time encoding requirement (i.e., the video's framerate) and hence is 50 frames per second. An encoding mode switch is considered at every segment while the results of every segment, as well as the average ones over the entire video sequence, are displayed.

The results of Table XI show that the minimum bitrate mode produced savings of 11% over static encoding, while the average PNSR also increased. From the results, we can see that the encoding rate was reduced to 49.33 frames per second, suggesting that our optimization took slightly longer than the static encoding. Still, the achieved encoding rate in FPS was very close to the imposed requirement. Bitrate savings in the particular scenario are attributed to the dense encoding space used to generate the forward prediction models that allow considering different encoding setups during the adaptation decision.

The next optimization scenario targeted two interlinked objectives while leveraging VMAF to monitor quality. First, to investigate whether substantial bitrate savings were possible while maintaining a VMAF level that is virtually indistinguishable from the static example of Table X. Second, to demonstrate the ability of our proposed approach to delivering high-quality video streams under extreme bandwidth fluctuations that can cause a dramatic bandwidth decrease and/ or scenarios where the available bandwidth is shared amongst users with equivalent quality of service plans. In such events, adaptive mechanisms such as the one proposed in this study need to be in place to secure the continuity and quality of the video streaming service.

In that context, for the example in Table XII, we reduced the VMAF constraint by three so that $92.58 - 3 = 89.58$. More generally, the 3-point VMAF reduction is also supported by the work on establishing just noticeable differences (JND) [37], [38]. The JND study described in [39] arrived at the conclusion that a 6-point VMAF difference is the threshold value after which the perceived difference in quality becomes noticeable between two compressed videos.

The results depicted in Table XII materialized both goals. In particular, a 32.8% reduction in original bitrate was achieved at only a slight reduction of the VMAF constraint (or the original PSNR). As expected, the final videos were indistinguishable from the original. These substantial savings were achieved at a frame rate of 58.63 frames per second, a rate that is well above the requirement of 50 frames per second.

*II. Minimum Bitrate Demands using SVT-AV1*

To investigate the universality of the proposed methods over different video codecs, we also considered SVT-AV1 for the minimum bitrate mode on a different video. Table XIII presents the average quality, bitrate, and encoding rate values for the Pedestrian video segments that serve as the baseline, static encoding constraints. We then followed the same approach as x265 in setting up the constraints. The results are presented in Tables XV and XVI.

For PSNR, the minimum bitrate optimization mode reduced bitrate requirements by 7% compared to the static approach of Table XIV. The gains primarily come from increasing the QP and involving a different encoding structure in the encoding setup of Seg. 2. In the opposite direction, the desired quality constraints are matched neither by PSNR nor by VMAF values besides the 1st segment. As already described, the latter is attributed to the tolerable percentage error that is introduced in order to avoid the ping-pong effect of switching between prediction models and thus

TABLE XIII
FORWARD PREDICTION MODELS COEFFICIENTS PER SEGMENT AND OPTIMIZATION OBJECTIVE – SVT-AV1 FOR PEDESTRIAN 432P VIDEO.

| | Segments | c = HL4ALT0 | | | | c = HL4ALT2 | | | | c = HL4ALT8 | | | |
|---|---|---|---|---|---|---|---|---|---|---|---|---|---|
| | | $\beta_{i,c,j,0}$ | $\beta_{i,c,j,1}$ | $\beta_{i,c,j,2}$ | $R^2$ | $\beta_{i,c,j,0}$ | $\beta_{i,c,j,1}$ | $\beta_{i,c,j,2}$ | $R^2$ | $\beta_{i,c,j,0}$ | $\beta_{i,c,j,1}$ | $\beta_{i,c,j,2}$ | $R^2$ |
| $\log(PSNR_i)$ | Seg 0 | 3.87 | -0.0015 | -6.64E-05 | 0.990 | 3.86 | -0.0013 | -6.74E-05 | 0.990 | 3.86 | -0.0013 | -6.71E-05 | 0.990 |
| | Seg 1 | 3.89 | -0.0033 | 0 | 0.989 | 3.88 | -0.0030 | 0 | 0.988 | 3.88 | -0.0031 | 0 | 0.987 |
| | Seg 2 | 3.89 | -0.0031 | 0 | 0.987 | 3.88 | -0.0030 | 0 | 0.987 | 3.89 | -0.0030 | 0 | 0.986 |
| $\log(VMAF_i)$ | Seg 0 | 4.45 | 0.0127 | -0.000264 | 0.980 | 4.45 | 0.0128 | -0.000265 | 0.980 | 4.45 | 0.0128 | -0.0003 | 0.990 |
| | Seg 1 | 4.48 | 0.0108 | -0.000200 | 0.994 | 4.48 | 0.0109 | -0.000200 | 0.994 | 4.48 | 0.0111 | -0.0002 | 0.994 |
| | Seg 2 | 4.49 | 0.0099 | -0.000200 | 0.992 | 4.49 | 0.0099 | -0.000200 | 0.992 | 4.49 | 0.0100 | -0.0002 | 0.991 |
| $\log(Bitrate_i)$ | Seg 0 | 8.82 | -0.0358 | -0.0003 | 0.99 | 8.77 | -3.38E-02 | -0.000325 | 0.99 | 8.69 | -0.0316 | -3.43E-04 | 0.98 |
| | Seg 1 | 8.91 | -0.0361 | -0.0003 | 1 | 8.87 | -0.0351 | -0.000300 | 1 | 8.78 | -0.0327 | -0.000300 | 1 |
| | Seg 2 | 8.79 | -0.0409 | -0.0003 | 1 | 8.73 | -0.0389 | -0.000300 | 1 | 8.62 | -0.0358 | -0.000300 | 1 |
| $\log(FPS_i)$ | Seg 0 | 3.46 | 0.0097 | 3.91E-05 | 0.980 | 3.47 | 0.0076 | 7.26E-05 | 0.990 | 3.44 | 0.0052 | 8.78E-05 | 0.990 |
| | Seg 1 | 3.46 | 0.0104 | 0 | 0.977 | 3.38 | 0.0126 | 0 | 0.974 | 3.36 | 0.0114 | 0 | 0.963 |
| | Seg 2 | 3.44 | 0.0126 | 0 | 0.97 | 3.42 | 0.0123 | 0 | 0.973 | 3.40 | 0.0104 | 0 | 0.977 |

$j \in \{PSNR, VMAF, BITRATE, FPS\}$

The $\beta_{i,c,j,0}$, $\beta_{i,c,j,1}$, and $\beta_{i,c,j,2}$ coefficients have been rounded up to 2, 4, and 6 significant digits, respectively.

$R^2$ refers to the Adjusted $R^2$. The adjusted $R^2$ values have been rounded to three significant digits.

compromising real-time performance by adding complexity. Nonetheless, the documented drop is within acceptable limits. Moreover, the encoding rate in terms of FPS is significantly higher than the minimum values required to achieve real-time performance.

In terms of the 3-point VMAF scenario leveraging the JND approach, results appear in Table XVI. The minimum bitrate optimization mode reduced bitrate requirements by approximately 50% while matching real-time encoding requirements. The VMAF score was reduced by 3.27; however, it remained well within the JND requirements and did not compromise perceptual quality at the displayed resolution.

## V. DISCUSSION AND CONCLUDING REMARKS

An adaptive video encoding methodology that is applicable to different video codecs for optimizing the utilization of available recourses is proposed. The approach uses multi-objective optimization to meet dominant, time-varying constraints in a video streaming session, namely video quality, bandwidth availability, and encoding rate.

Results demonstrate that the proposed methods can effectively mitigate time-varying bandwidth fluctuations that are likely to result in buffering incidents and thus degraded end user's QoE. The applicability across the range of examined video codecs holds great promise, especially in view of the fast-changing landscape of the video compression industry highlighted in this study. In that context, following VVC/H.266 standardization, MPEG has subsequently released two new video codecs, termed Essential Video Coding (EVC) [40] and Low-complexity Enhancement Video Coding (LCEVC) [40]. The goal is to alleviate patent/ licensing schemes on the one hand and complexity concerns on the other. As a result, approaches such as the ones described in this study are now, more than ever, timely and central to the efficient utilization of individual codecs to benefit the user experience.

Our ongoing research is focused on interfacing with a wireless network simulator to validate the proposed methods under realistic unicast and multicast video transmission scenarios for both live and on-demand sessions. Furthermore, we are also working on extending the multi-objective optimization framework to include decoding time and decoding power consumption, and adopting blind VQA metrics at the receiver, to drive the adaptive video encoding process.

TABLE XIV
STATIC MODE (DEFAULT) – YOUTUBE RECOMMENDED SETTINGS – SVT-AV1 – PEDESTRIAN 432P VIDEO

| Seg.ID | CQP | GOP | FIL | Bitrate (kbps) | FPS | PSNR (dB) | VMAF |
|---|---|---|---|---|---|---|---|
| Seg. 0 | 20 | HL4ALT8 | OFF | 2,429 | 37.44 | 45.19 | 99.67 |
| Seg. 1 | 20 | HL4ALT8 | OFF | 2,492 | 37.67 | 44.55 | 99.81 |
| Seg. 2 | 20 | HL4ALT8 | OFF | 2,042 | 39.12 | 45.03 | 99.21 |
| Avg | 20 | | | 2,321 | 38.08 | 44.92 | 99.56 |

[1]Filters=OFF

TABLE XV
SVT-AV1 MINIMUM BITRATE OPTIMIZATION MODE - PEDESTRIAN 432P VIDEO[1].

| Enc. Segs. | GOP | QP | Bitrate (kbps) | FPS | PSNR | VMAF |
|---|---|---|---|---|---|---|
| Seg. 0 | HL4ALT0 | 26 | 2,191 | 41.66 | 44.98 | 99.48 |
| | | constraints | | ≥25 | ≥44.92 | |
| Seg. 1 | HL4ALT0 | 26 | 2,348 | 42.27 | 44.87 | 99.67 |
| | | constraints | | ≥25 | ≥44.92 | |
| Seg. 2 | HL4ALT2 | 26 | 1,878 | 43.93 | 44.20 | 98.81 |
| | | constraints | | ≥25 | ≥44.92 | |
| Avg | | | 2,139 | 42.62 | 44.68 | 99.32 |
| ΔGains | | | 7.8% | 4.54 | (-0.24) | (-0.24) |

[1]Predicted values using the models depicted in Table XI.

TABLE XVI
SVT-AV1 MINIMUM BITRATE OPTIMIZATION MODE - PEDESTRIAN 432P VIDEO[1]. JUST NOTICEABLE DIFFERENCE 3-POINT VMAF EXAMPLE.

| Enc. Segs. | GOP | QP | Bitrate (kbps) | FPS | PSNR | VMAF |
|---|---|---|---|---|---|---|
| Seg. 0 | HL4ALT8 | 37 | 1,162 | 46.15 | 41.54 | 96.70 |
| | | Constraints | | ≥25 | | ≥96.56 |
| Seg. 1 | HL4ALT2 | 37 | 1,247 | 47.31 | 41.11 | 96.23 |
| | | constraints | | ≥25 | | ≥96.56 |
| Seg. 2 | HL4ALT2 | 38 | 1,086 | 46.01 | 42.03 | 95.94 |
| | | constraints | | ≥25 | | ≥96.56 |
| Avg | | | 1,165 | 46.49 | 41.56 | 96.29 |
| ΔGains | | | 49.8% | 8.41 | (-3.36) | (-3.27) |

[1]Predicted values using the models depicted in Table XI.

## ACKNOWLEDGMENT

The authors would like to Krishna A. Poddar for his valuable contribution in preparing the results tables.